\documentclass[%
 reprint,
 amsmath,amssymb,
 aps,
 prl,
floatfix,
]{revtex4-1}

\usepackage{graphicx}% Include figure files
\usepackage{dcolumn}% Align table columns on decimal point
\usepackage{bm}% bold math
\usepackage{color} %colored text
\begin{document}
\title{DNA-coated Functional Oil Droplets}% Force line breaks with \\

\author{Alessio Caciagli}
\author{Mykolas Zupkauskas}
\author{Thomas O'Neill}
\affiliation{Optoelectronics, Cavendish Laboratory, University of Cambridge, Cambridge CB3 0HE, United Kingdom}
\author{Aviad Levin}
\affiliation{Department of Chemistry, University of Cambridge, Cambridge CB2 1EW, United Kingdom}
\author{Tuomas P. J. Knowles}
\affiliation{Department of Chemistry, University of Cambridge, Cambridge CB2 1EW, United Kingdom}
\affiliation{Cavendish Laboratory, University of Cambridge, Cambridge CB3 0HE, United Kingdom}
\author{Cl\`ement Mugemana}
\author{Nico Bruns}
\affiliation{Adolphe Merkle Institute Chemin des Verdiers 4 CH-1700 Fribourg Switzerland}
\author{William J. Frith}
\affiliation{Unilever R\&D Colworth, Colworth Science Park, Sharnbrook, Bedfordshire, MK44 1LQ, United Kingdom}
\author{Erika Eiser}\email{ee247@cam.ac.uk}
\affiliation{Optoelectronics, Cavendish Laboratory, University of Cambridge, Cambridge CB3 0HE, United Kingdom}

\date{\today}% It is always \today, today,

\begin{abstract}
Many industrial soft materials often include oil-in-water (O/W) emulsions at the core of their formulations. By using tuneable interface stabilizing agents, such emulsions can self-assemble into complex structures. DNA has been used for decades as a thermoresponsive highly specific binding agent between hard and, recently, soft colloids. Up until now, emulsion droplets functionalized with DNA had relatively low coating densities and were expensive to scale up. Here a general O/W DNA-coating method using functional non-ionic amphiphilic block copolymers, both diblock and triblock, is presented. The hydrophilic polyethylene glycol ends of the surfactants are functionalized with azides, allowing for efficient, dense and controlled coupling of dibenzocyclooctane functionalized DNA to the polymers through a strain-promoted alkyne-azide click reaction. The protocol is readily scalable due to the triblock’s commercial availability. Different production methods (ultrasonication, microfluidics and membrane emulsification) are used with different oils (hexadecane and silicone oil) to produce functional droplets in various size ranges (sub-micron, $\sim 20\,\mathrm{\mu m}$ and $> 50\,\mathrm{\mu m}$), showcasing the generality of the protocol. Thermoreversible sub-micron emulsion gels, hierarchical ``raspberry'' droplets and controlled droplet release from a flat DNA-coated surface are demonstrated. The emulsion stability and polydispersity is evaluated using dynamic light scattering and optical microscopy. The generality and simplicity of the method opens up new applications in soft matter and biotechnological research and industrial advances.
\end{abstract}
\maketitle

Oil-in-water (O/W) emulsions are biphasic liquid materials with critical importance in a broad range of industrial fields such as food science, personal care, pharmaceuticals and paints \cite{Tadros2013}, and as templates for advanced materials fabrication \cite{Binks2006,Manoharan2003}. If the surfactant or particle employed to stabilize the interface is responsive to external stimuli such as pH \cite{Macon2015}, temperature \cite{Tsuji2008} or light \cite{Zarzar2015}, O/W emulsions can be effectively used as reconfigurable fluids. The means to promote their self-organization into complex architectures in a programmable way is a rapidly developing research topic as it can potentially lead to a new class of dynamically tuneable soft materials. \\
DNA hybridization has already proved to be ideal for the creation of functional and patchy particles. Due to its thermo-reversibility and high specificity, it has been extensively employed to drive and direct colloidal self-assembly via temperature control \cite{Feng2013,Jones2010,Macfarlane2011}. Upon choosing a suitable combination of surfactants and ligands, O/W emulsions \cite{Hadorn2012,Feng2013a,Joshi2016} and vesicles \cite{Parolini2015,Hadorn2016} have been successfully functionalized with DNA. The unique property of these objects is their natural ``patchiness'' due to the mobility of the linkers on the interface, which in turn allows for programmed sequential self-assembly of the building blocks \cite{Zhang2017}. The established the potential for using DNA-functionalized O/W emulsions in materials applications. However, currently available protocols either rely on cholesterol anchors or on avidin-biotin chemistry \cite{Diamandis1991,Dreyfus2009}. When coupled to emulsion preparation, these methods relyeither on a limited surfactant choice and cumbersome fabrication \cite{Joshi2016}, or on expensive components \cite{Feng2013,Zhang2017} and only offer a limited control over the coating density of the oligonucleotides on the liquid interface. This poses a severe limitation to the versatility of the system and its scalability. \\
Here we develop a simple and general protocol to produce DNA-functionalized O/W emulsions. The method is suitable for any emulsification strategy, can be used with different oil families (hydrocarbons and silicone oils) and can also be pursued using commercially available materials, making the procedure potentially scalable. Di- and tri-block copolymers with hydrophilic chain(s) of polyethylene oxide (PEG) are employed as non-ionic surfactants, stabilizing the droplets. These classes of surfactants offer some advantages over ionic surfactants such as sodium dodecyl sulphate (SDS): They are less sensitive to the presence of electrolytes \cite{Schick1987}, allow for easier formulations due to the ability to systematically tune the emulsifiers’ polarity \cite{Schick1987} and offer enhanced emulsion stability to the interfacial film between approaching droplets, preventing their coalescence \cite{Kong2010}. Moreover, these non-ionic block copolymers show a much higher chemical stability. In our system, the block copolymer’s free end(s) are functionalized with an azide group (N$_3$) (see Experimental Section). To stress again the generality of our method, both diblock and triblock (PEO-PPO-PEO; with the middle block being polypropylene oxide) copolymers have been employed in the study. The choice of the former or latter depends on the desired emulsion properties such as size distribution, stability, or DNA coating density, but it is irrelevant to the success of the protocol. We used 24 bases long single-stranded (ss)DNA strands (purchased from IDT) that were similar to those reported in Zupkauskas et al. \cite{Zupkauskas2017}. The ssDNA, called \textbf{A} DNA, starts with an amine-functionalized 5' end, and consists of a 15 thymine (T) long spacer followed by a 9 bases long single stranded ``sticky end''. The same structure was used for the ssDNA \textbf{A'} holding the complementary sticky end. \textbf{A} and \textbf{A'} strands were mixed with dibenzocyclooctyne-sulfo-N-hydroxysuccinimidyl ester (DBCO-sulfo-NHS) in separate containers, resulting in DBCO-ssDNA strands, which were reacted to the azide terminated chains (detailed reaction conditions are given in the Experimental Section).

\begin{figure*}[t]
\includegraphics[width=0.8\textwidth]{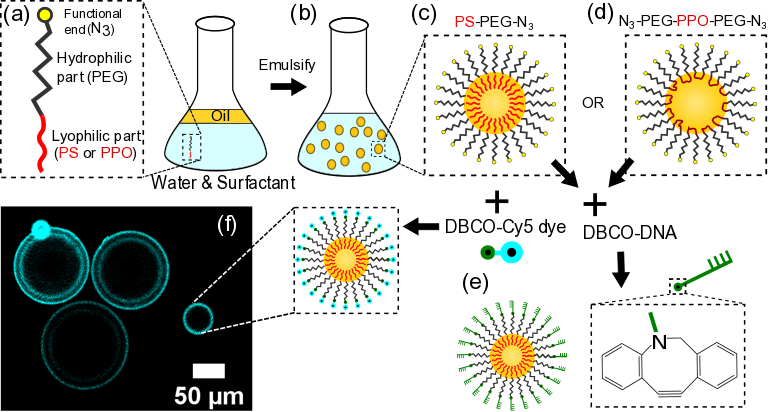}
\caption{Schematic protocol of making DNA-functionalized oil droplets. (a) An aqueous solution containing the bock-copolymer PS-PEG or the triblock copolymer PEG-PPO-PEG with azide-functionalized PEG-ends is mixed with immiscible silicone- or hexadecane oil. (b) After emulsification through microfluidics, membrane or sonication we obtain oil-in-water droplets stabilized by a densely packed PEG-corona ((c) or (d)). Using click-chemistry, the functionalized PEG ends are then reacted either to the fluorescently labelled DBCO-Cy5 dye and imaged in confocal microscopy (f) or alternatively to DBCO-DNA (e). Note that the second inner ring visible in the confocal image is due to lensing and is not a signature of a double emulsion.}
\end{figure*} 
The procedure for droplet formation and functionalization is sketched in Figure 1. First, the azide-functionalized block copolymer (Figure 1a) was dispersed in de-ionized (DI) water - emulsification was achieved according to one of the three chosen strategies detailed later (Figure 1 and Table 1). In the resulting droplets the hydrophobic block of the polymer acted as anchor in the oil phase, while the hydrophilic PEG-N$_3$ segment stretched into the water phase forming a sterically stabilizing dense brush (Figure 1c, d). Lastly, the single DBCO-ssDNA strands were attached to the N3 ends of the block copolymers via a strain-promoted alkyne-azide click reaction (SPAAC) (Figure 1d). The success of this functionalization strategy was demonstrated by confocal microscopy on emulsion droplets for which the DBCO-DNA strands were replaced by fluorescently labelled DBCO (Figure 1e). The DNA coupling mechanism employed here has been shown to deliver colloidal particles with very high DNA-coating densities \cite{Oh2015}. This is possible thanks to the high yield of the SPAAC reaction (typically larger than 90\%) and the small size of the grafting point in comparison with the avidin-biotin chemistry used in previous works \cite{Geerts2010}. Indeed, the polymer-DNA grafting density is expected to be enhanced in our DNA-functionalized emulsions since the packing of the block copolymer surfactant on the fluid emulsion-droplet interface is typically very tight and its surface density is at least an order of magnitude larger than that of the streptavidin used in previous studies \cite{Joshi2016,Zhang2017}.\\
The emulsion droplets were produced by three different emulsification methods: membrane emulsification \cite{Charcosset2004,Egidi2008}, microfluidics and ultrasonication \cite{Canselier2002}. The first two methods produce so-called macro-emulsions (droplet size larger than $1\,\mathrm{\mu m}$) differing in volume throughputs \cite{Shah2008,Vladisavljevic2012} while the third produces typically nano-emulsions (droplet size on the order of a few $100\,\mathrm{nm}$ \cite{Solans2005,Gupta2016}). Together, they span a wide range of emulsion characteristics and sizes, which constitutes a good benchmark for the generality of our protocol. For each method, two combinations of oil and surfactant were chosen: hexadecane with a custom PS-PEG diblock copolymer \cite{Zupkauskas2017}, and silicone oil with a commercial triblock copolymer (Synperonic F108). The results are summarized in Table 1. 
\begin{table}[t]
\includegraphics[width=0.45\textwidth]{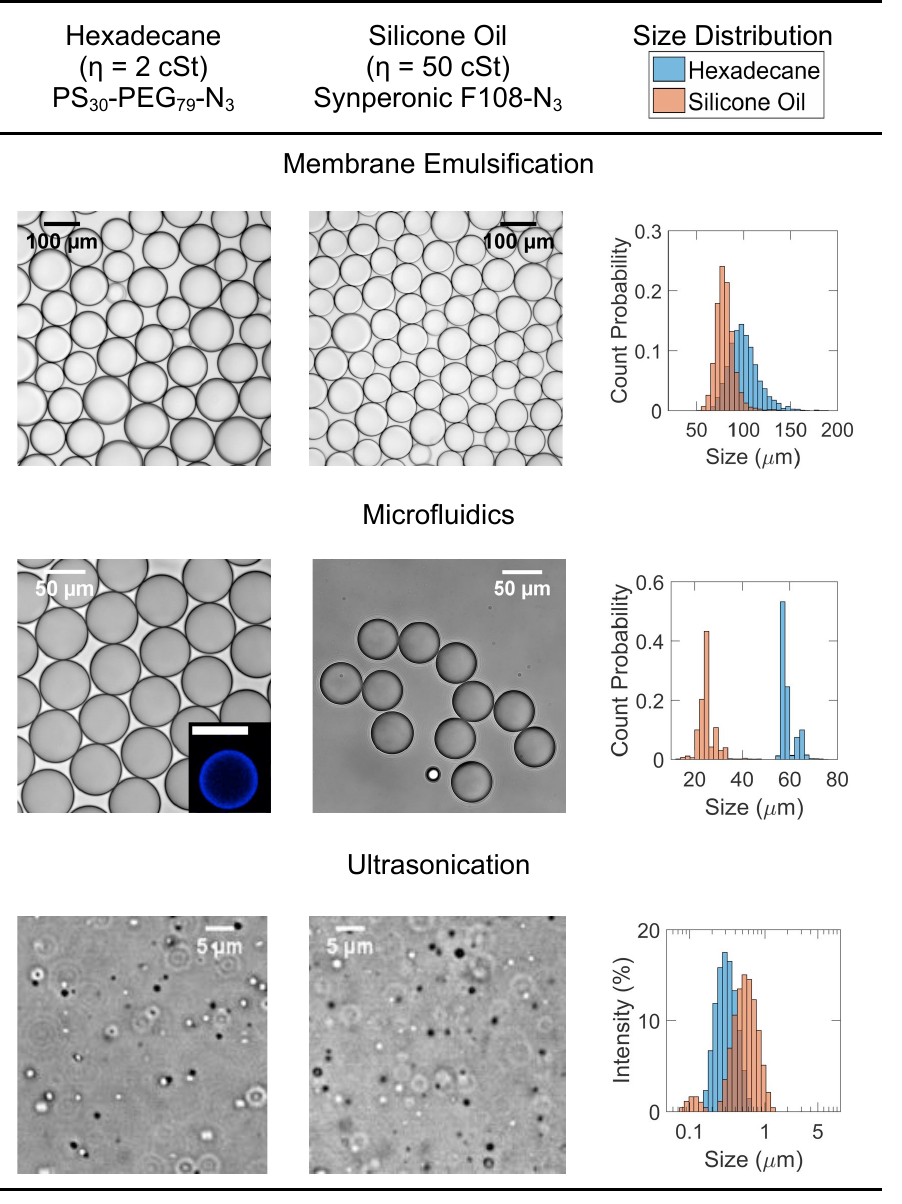}
\caption{Comparison of functional droplets produced by different emulsification methods. The images are obtained by brightfield optical microscopy. The inset in the hexadecane column, microfluidics row is obtained by confocal microscopy. The size distributions for membrane emulsification and microfluidics are obtained by image analysis, whereas that for ultrasonication is obtained by Dynamic Light Scattering (DLS).}
\end{table} 
The membrane emulsification strategy produced a slightlypolydisperse emulsion according to the method’s typical yield \cite{Egidi2008}. A lower polydispersity was achieved using the triblock copolymer/silicone oil combination. Rather than an effect of the surfactant, this improved monodispersity was due to the higher viscosity ratio between dispersed and continuous phase, which decreased the supply of the dispersed phase during droplet formation, resulting in smaller droplets \cite{Kukizaki2009,Lloyd2014}.
The droplets have been shown to be stable against coalescence for months, with a slight broadening of the size distribution over time due to Ostwald Ripening. Droplets produced by microfluidics were very monodisperse with polydispersity index (PDI) of less than $\sim 10\%$. The diblock copolymer/hexadecane combination produces larger droplets (probably as a result of the lower viscosity ratio between the dispersed and continuous phases) and of slightly better quality, due to faster kinetics of surfactant adsorption at the interface. We ascribe this fast kinetics to two effects. First, the adsorption is faster for lower viscosity ratios between the dispersed and continuous phase, which favours the hexadecane system. Secondly, although triblock copolymers have a greater interfacial anchoring strength than diblock copolymers of the same geometry, their diffusion is slower due to their larger size \cite{Holtze2008}. This again favours the hexadecane system. For both systems, however, droplet stability is also very good and provides a longer life time and resistance to coalescence at higher temperatures. The ultrasonication emulsification strategy produced a more polydisperse nano-emulsion, which is typical for the method \cite{Solans2005}. While it has not been attempted, a more narrow distribution might be achievable when fine tuning the overall procedure \cite{Leong2009,Gaikwad2008}. The diblock copolymer/hexadecane combination gave again superior quality droplets (smaller and more monodisperse), which can be explained using the same arguments given in the microfluidics case. As for the membrane emulsification droplets, the emulsions show a slight broadening of the size distribution over time due to Ostwald Ripening. Summarising, for all the three emulsification methods studied we show an excellent stability of our emulsions against coalescence. Membrane emulsification and ultrasonication droplets show a slight broadening of the size distribution due to Ostwald Ripening over long shelf times, though the rate is relatively slow. While both oil/surfactant combinations produce excellent droplet quality, the diblock/hexadecane combination gives a better PDI: we ascribe this to a more favourable surfactant/oil interaction (silicone oils are known to require silicone surfactants for best performance \cite{OLenick2000}).

\begin{figure*}[t]
\includegraphics[width=0.8\textwidth]{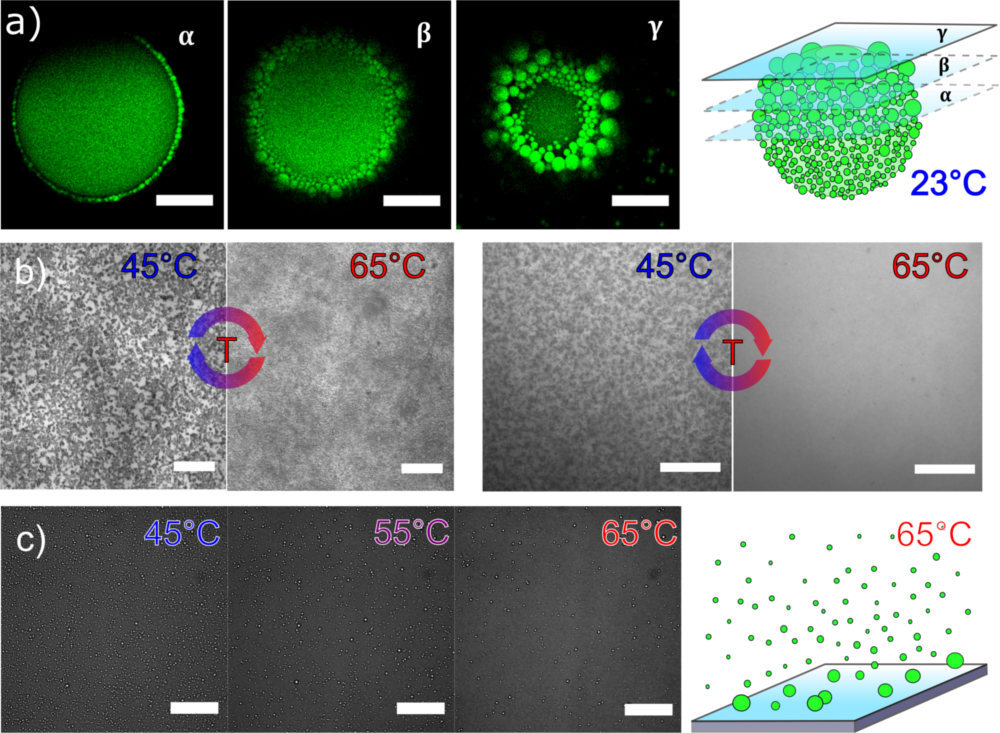}
\caption{ (a) Three confocal images at different heights, $\alpha$, $\beta$ and $\gamma$, of a roughly 60 $\mathrm{\mu m}$ large, membrane emulsified silicone-oil droplet coated with \textbf{A}-PEG-PPO-PEG-\textbf{A} DNA coating
(measured at room temperature). These are covered by ultrasonically (US) emulsified
hexadecane droplets coated with PS-PEG-\textbf{A'} DNA, as sketched right. (b, left) A thermal
cycle of ultrasonicated droplets (half with PS-PEG-\textbf{A}, half with PS-PEG-\textbf{A'}) showing an
emulsion gel below and a gas of emulsion droplets above $T_m$. (right) Corresponding
measurement of 420 nm large PS colloids, half coated with \textbf{A}-PEG-PPO-PEG-\textbf{A} and the other
with \textbf{A'}-PEG-PPO-PEG-\textbf{A'} DNA. (c) Brightfield image of a flat, \textbf{A} DNA coated glass surface
to which ultrasonicated hexadecane droplets with PS-PEG-\textbf{A'} DNA are hybridized at $T$ =
45 $^{\circ}$C. With increasing temperature the droplets melt off progressively with the smaller
droplets being released first. Scale bar = 20 $\mathrm{\mu m}$.}
\end{figure*} 
To show the possibility of using the oil droplets (ODs) as microscale building blocks we present three different experiments. In the first we constructed ``raspberry droplets'' (Figure 2a). The $60\,\mathrm{\mu m}$ large ``core'' silicone-oil droplets were prepared in a membrane emulsifier and stabilized by \textbf{A}-PEG-PPO-PEG-\textbf{A} DNA. The small droplets were made with the ultrasonic probe and functionalized with the complementary \textbf{A'} DNA (hexadecane PS-PEG-\textbf{A'}); the latter had a range of sizes from about $200\,\mathrm{nm}$ to $5\,\mathrm{\mu m}$. We added a small amount of BODIPY fluorescent dye to these hexadecane droplets. After mixing the two populations of droplets in phosphate buffer in a flat capillary and incubating the samples at room temperature for 15 minutes, which is well below the melting temperature for the \textbf{AA'} bond used here, we performed confocal microscopy. The recoded images, taken at different heights of the larger ODs as shown in Figure 2a, revealed raspberry-like composite droplets, with the small drops anchored to the large ones owing to the \textbf{AA'} DNA hybridization. Note, utilising the specific \textbf{AA'} DNA binding prevents the small and large droplets to bind to each other at all temperatures between $5\,^{\circ} \mathrm{C}$ and $90\,^{\circ}\mathrm{C}$, as \textbf{AA} or \textbf{A'A'} bonds are forbidden. More droplets were found at the top of the big drops due to buoyant forces caused by hexadecane's density that is lower than that of the aqueous buffer solution. Moreover, we noticed that although the dense PEG-DNA-PEG sandwich between the small and large ODs prevents them from coalescing the initially fluorophore-free large OD became also fluorescently active. This suggests that the hydrophobic BODIPY dye was able to diffuse across the surfactant barrier. We argue that this is due to osmotic-pressure driven Oswald ripening. \\
In a second approach, two batches of ultrasonicated, hexadecane droplets (one stabilized via PS-PEG-\textbf{A} DNA and the other via PS-PEG-\textbf{A'} DNA) were mixed in a flat capillary at a total oil-concentration of 5\% v/v and then imaged. The sample was heated to $70\,^{\circ}\mathrm{C}$, above the melt temperature of the \textbf{AA'} bond ($T_\mathrm{m} = 65\,^{\circ}\mathrm{C}$), held there for 15 minutes and then cooled down to the two-phase region in which the complementary DNA strands start to bind. We used a cooling rate of $1\,{\circ}\mathrm{C\,min^{-1}}$ to ensure equilibrium DNA hybridization. The resulting droplet aggregates, shown in Figure 2b left, resemble those of gels made of DNA-coated colloids of similarly sized hard polystyrene particles (Figure 2b, right), however, these colloidal gels displayed a lower melting temperature ($\sim 60\,^{\circ}\mathrm{C}$, respectively \cite{Zupkauskas2017}. In order to understand this difference in $T_\mathrm{m}$ we need to mention that the melting transition between colloids densely coated complementary ssDNA is as sharp as $1-2\,^{\circ}\mathrm{C}$ \cite{Geerts2010}. This is due to the fact that several \textbf{AA'} pairs can form in the contact region. $T_\mathrm{m}$ increases logarithmically with an increasing total number of possible hydrogen bond that can form between the ssDNA pairs and sharpens as well \cite{Geerts2010}. Hence a possible reason for the difference in $T_\mathrm{m}$ between colloidal gels and emulsion droplets is the mobility of the ssDNA on the OD interface that allows for an even tighter rearrangement of DNA in the contact region once two droplets meet. Further, the gain in binding energy also allows the soft ODs to flatten, thus leading to a slightly increased effective size of the contact patch and therefore a higher binding strength – this increase is only limited by the cost to deform the interface \cite{Feng2013a}. Note, the binding energy of a single \textbf{AA'} pair is about $16\,k_\mathrm{B}T$ ($k_\mathrm{B}$ being the Boltzmann constant) \cite{Zupkauskas2017}. Hard polystyrene colloids do not allow for a surface deformation and DNA attached to their surface are not mobile. These ``emulsion gels'' were completely thermally reversible over 5 heating-cooling cycles, showing a narrow melting region of $\Delta T \sim 2\,^{\circ}\mathrm{C}$ and a sharp melt temperature, similar to the PS-colloid gel. Remarkably, no measurable coalescence was observed. \\
In a third experiment we demonstrate that the same hexadecane droplets can be DNA-anchored to a flat surface and then released in a controlled manor. As a number of chemical and pharmaceutical compounds are soluble in oil but insoluble in water, this system is a good model for those applications. Here the surface of a flat capillary was first coated with \textbf{A} DNA using a grafting method via the comb-polymer polylysine-PEG-biotin \cite{Yanagishima2012,Geerts2010}, which adsorbs to the plasma treated glass surfaces with the positively charged polylysine backbone. Streptavidin was then used to bind biotinylated \textbf{A} DNA to polylysine-PEG-biotin on the surface. After flushing the capillary 5 times with pure buffer solution, it was injected with a 1\% v/v solution containing PS-PEG \textbf{A'}-functionalized US-hexadecane droplets at room temperature. The sample was then heated to $70\,^{\circ}\mathrm{C}$, inverted and cooled down to coat the bottom surface with droplets such that imaging with our inverted microscope was possible. Imaging the sample in an optical microscope showed how the buoyant ODs detached and rose upon heating (Figure 2c). Although we observed a clear release of the ODs from the surface, the melting region was larger than that for hard-colloidal gels simply because of the much larger size-distribution of the ODs: The smaller ODs were released first as their effective contact area with the flat support surface was smaller, which is visible in the microscope images in Figure 2c.

In conclusion, we have demonstrated a simple method to make oil-in-water emulsion droplets functionalized with DNA.  The method is applicable to virtually all emulsification strategies since the functionalization is carried directly on the surfactant. It is furthermore suitable for any non-ionic surfactant containing a hydrophilic PEO chain and any oil that can be emulsified by this class of surfactants (or a mixture containing at least one of them), making it very scalable. The self-organization properties conferred by the specific, DNA-mediated interaction are demonstrated by creating hierarchical droplet structures (``raspberry droplets'') and a thermo-responsive emulsion gel. Furthermore, our DNA-functionalized OD formulation represents a simple surface-controlled release system.\\
These properties open new routes toward the realization of hierarchic self-assembly of micro-reactors that can be loaded with various chemical or other compounds such as nano-particles and polymers \cite{Tkachenko2011,Patra2017}. Such compartmentalized formulations will be of great interest in prolonging the shelf-life of emulsions relevant in pharmaceuticals and foods. In contrast to patchy colloidal particles with static linkers, the mobile linkers on the emulsion droplets ensure progress toward the thermodynamic equilibrium of the self-assembly process \cite{Angioletti-Uberti2016}. Furthermore, the scalability and simplicity of the protocol ensures batch production of colloids with specific valency \cite{Wang2012} and represent a huge step towards the bulk realization of “colloidal molecules”. By careful design of the DNA-mediated interactions between droplets, emulsion gels with exotic properties could be realized such as fluids which harden upon heating \cite{Roldan-Vargas2013} or with tunable porous morphologies conferring novel optical properties \cite{Pine1997}. Coupling the compartmentalization capability of emulsions with the sensitivity to environmental stimuli such as temperature, our system could be used as a highly efficient micro reactor or as an advanced drug and cargo delivery system, with broad applications in soft matter, medical and biotechnological research.

\section{Experimental Section}
\textit{Diblock copolymer synthesis:} The polystyrene-\textit{block}-PEG-azide (PS-b-PEG-N$_3$) was synthesized by RAFT (Reversible Addition Fragmentation chain Transfer) polymerization process in presence of azide-(PEG)-chain transfer agent. The latter was made by esterification between a RAFT initiator (4-cyano-4-[(dodecyl-sulfanylthiocarbonyl) sulfanyl]pentanoic acid) and a commercially available $\alpha$-Hydroxy-$\omega$-azido polyethylene glycol. The synthesized PEG chain transfer agent was then copolymerized with styrene yield the PS$_30$-\textit{b}-PEG-N$_3$ diblock copolymer. The synthesized azido-functionalized block copolymers were purified by precipitation in diethyl ether and further characterized by $^1$H-NMR, SEC and FTIR analysis.\\
\textit {Synperonic F108 functionalization with azide groups:} 1 g Synperonic F108 (PEG-PPO-PEG,
SigmaAldrich) was dissolved in 13 mL of dichloromethane (DCM) in a round-bottom flask.
420 $\mathrm{\mu L}$ triethylamine (TEA) was added and the flask was submerged into an ice-water mixture with the cap closed. 570 mg 4-toluenesulfonyl chloride (TsCl, SigmaAldrich) was
dissolved in 7 mL DCM. The solution was added to the cooled F108 and then stirred overnight
at 600 rpm, letting the ice-water mixture reach room temperature in ~2 h. After reacting, the
solution was evaporated under vacuum. The precipitate was washed 2 times with 3\% HCl in
MeOH and 3 times with MeOH, each time precipitated with diethyl ether at subzero temperatures. The F108-TsCl was dried under vacuum and then dissolved in dimethylformamide (DMF) containing 100 mg NaN$_3$ in a round-bottom flask. The solution was stirred at 65 $^{\circ}$C at 1000 rpm overnight. The purification was repeated as described above. The dry F108-N$_3$ was stored in the freezer.\\
\textit{DNA preparation:} The DNA strands were purchased from Integrated DNA Technologies
(IDT): amine-5’- TTT TTT TTT TTT TTT GGT GCT GCG-3’ (called \textbf{A}), amine-5’-TTT TTT TTT TTT TTT CGC AGC ACC-3’ (\textbf{A’}: sticky end complementary to \textbf{A}). Amine to 
dibenzylcyclooctane (DBCO) functionalization was done as described by Zupkauskas et al.\cite{Zupkauskas2017} The DBCO-DNA was kept frozen in 10 mM phosphate buffer (PB) at 0.05 mmol L\textsuperscript{-1}.\\
\textit{Membrane emulsification:} The emulsions were obtained by a membrane technique using a
LDC-1 Dispersion Cell (Micropore Technologies Ltd, Loughborough, UK). 5 mL of the dispersed phase (either hexadecane or silicone oil) was injected through the membrane (20 $\mathrm{\mu m}$ pore size, 80 $\mathrm{\mu m}$ intra-pore distance) into 50 mL of Millipore water containing 2\% w/v PS-PEG-N$_3$ or F108-N$_3$ by means of a syringe pump (model Aladdin 1000, WPI, Sarasota,
USA) with a flow rate of 0.5 mL min\textsuperscript{-1}. The agitator was driven by a 24V DC motor and the paddle rotation speed was set to 19.066 Hz (1144 rpm) corresponding to an applied voltage of 10V.\\
\textit{Ultrasonic emulsification:} 1600 $\mathrm{\mu L}$ F108-N$_3$ (or PS-PEG-N$_3$) at 2\% w/v in Millipore water
was mixed with 200 $\mathrm{\mu L}$ hexadecane and 200 $\mathrm{\mu L}$ water in a glass vial. The mixture was prevortexed and ultrasonicated with a probe (Bandelin Sonopuls HD 2200) at 20\% amplitude for 10 minutes in pulsed mode (0.5 s on, 0.5 s off). The resulting block-copolymer-azide functionalized droplets were kept at room temperature. \\
\textit{Microfluidics:} The oil phase (hexadecane or polydimethylsiloxane, PDMS) was flowed at 50
$\mathrm{\mu L}$ h\textsuperscript{-1} and the water phase containing 2\% w/w PS-PEG-N$_3$ or F108-N$_3$ was flowed at 250 $\mathrm{\mu L}$ h\textsuperscript{-1} in a T-junction microfluidic device. The droplets were kept at room temperature. \\
\textit{General DNA attachment protocol:} The azide-functionalized droplets were washed 4-5 times with 0.5\% w/w Synperonic F108 dissolved in PB. For sub-micron sized droplets, 100 $\mathrm{\mu L}$ of drops at ~70\% v/v were mixed with 25 nanomoles of DBCO-DNA in a total volume of 1 mL PB containing 0.5\% F108 (A and A’ reacted separately). The mixture was gently shaken at
room temperature for 24 h, while raising the NaCl concentration to 100 mM to aid the reaction (screen DNA charge) incrementally over the first 6 hours. The droplets were then washed 5 times with 0.5\% w/w F108 in PB and kept at room temperature. For larger (5 $\mathrm{\mu m}$ and up) droplets, 10 times less DNA was used for the same volume of droplets. DNA grafting to flat surfaces was done following the protocol by Yanagishima et al.\cite{Yanagishima2012}\\
\textit{Sample preparation:} All samples with DNA contained 10 mM PB with 50 mM added NaCl to
ensure good DNA hybridization.\\
\textit{Imaging and characterization:} Optical microscopy was done using a Nikon Eclipse inverted microscope with dry 20x 0.75 NA and 40x 0.95 NA objectives. Confocal microscopy was done with a Leica TCS SP5 microscope using a 63x oil immersion objective. Dynamic light scattering was done using a Malvern Zetasizer ZS. The fluorescence assay was done with a fluorimeter using a DBCO-Cy5 dye (Life Sciences)

\textbf{Acknowledgements}
AC and MZ contributed equally to this work. MZ would like to thank EPSRC and Unilever for the CASE award RG748000. AC and EE acknowledge the ETN-COLLDENSE (H2020-MCSA-ITN-2014, Grant No. 642774) and the Winton Program for the Physics of Sustainability. AL and TK acknowledge the European Research Council under the European Union’s Seventh Framework Programme (FP7/2007 2013) through the ERC grant PhysProt (agreement No337969). NB and CM acknowledge financial support by the Swiss National Science Foundation (project PP00P2 144697 and National Centre of Competence in Research (NCCR) Bio-Inspired Materials) and by the KTI/CTI.

\bibliography{Drpolet_paper}
\end{document}